# Energy Efficient RAN Slicing and Beams Selection for Multiplexing of Heterogeneous Services in 5G mmWave Networks

PraveenKumar Korrai, *Member, IEEE*, Eva Lagunas, *Senior Member, IEEE*, Shree Krishna Sharma, *Senior Member, IEEE*, Symeon Chatzinotas, *Senior Member, IEEE*

*Abstract*—In this paper, we study a RAN resource-slicing problem for energy-efficient communication in an orthogonal frequency division multiple access (OFDMA) based millimeter-wave (mmWave) downlink (DL) network consisting of enhanced mobile broadband (eMBB) and ultra-reliable low-latency communication (URLLC) services. Specifically, assuming a fixed set of predefined beams, we address an energy efficiency (EE) maximization problem to obtain the optimal beam selection, Resource Block (RB), and transmit power allocation policy to serve URLLC and eMBB users on the same physical radio resources. The considered problem is formulated as a mixed-integer non-linear fractional programming (MINLFP) problem that considers minimum data rate and latency in packet delivery constraints. By leveraging the properties of fractional programming theory, we first transform the formulated non-convex optimization problem in fractional form into a tractable subtractive form. Subsequently, we solve the transformed problem using a two-loop iterative algorithm. The main resource-slicing problem is solved in the inner loop utilizing the difference of convex (DC) programming and successive convex approximation (SCA) techniques. Subsequently, the outer loop is solved using the Dinkelbach method to acquire an improved solution in every iteration until it converges. Our simulation results illustrate the performance gains of the proposed methodology with respect to baseline algorithms with the fixed and mixed resource grid models.

## I. INTRODUCTION

THE next generation of mobile networks is expected to enable a smart and connected community while supporting heterogeneous services including enhanced mobile broadband (eMBB), ultra-reliable low-latency communications (URLLC), and massive machine-type communications (mMTC) services [1]. The eMBB requires higher throughput (i.e., 1 Gbps or more) to support applications like ultra high definition (UHD) video streaming and virtual reality, while URLLC needs to achieve lower latency (i.e., even less than 1ms for some applications) in delivering packets to support applications such as telesurgery, vehicle to vehicle (V2V) communications, and factory automation.

To achieve all these stringent and sometimes conflicting requirements, the fifth generation (5G) of networks will include both an evolution of the conventional fourth-generation

The authors are with the SnT, University of Luxembourg, L-1855 Luxembourg City, Luxembourg (e-mail: {praveen.korrai, eva.lagunas, shree.sharma, symeon.chatzinotas}@uni.lu).

long-term evolution (4G-LTE) systems and the extension of a new radio access technology, standardized globally by 3GPP as new radio (NR) [2]. However, the 5G NR requires a large transmission bandwidth to support the above-mentioned services and their associated applications. In this context, the millimeter Wave (mm-Wave) frequency spectrum ranging from 28 GHz-300 GHz has been considered an enabler for 5G and beyond wireless networks [3]. These frequencies offer higher transmission bandwidths as compared to the ones below sub-6 GHz bands. Initial capacity calculations have revealed that networks operating at mm-Wave frequencies achieve significantly higher data rates than 4G networks [4]. In addition to a large amount of available bandwidth, the mm-Wave signal has a smaller wavelength than the conventional microwave signal, which gives the advantage of placing a high number of antennas in less area. As a result, considerable beam gains can be achieved to subside the occurred propagation losses in mm-Wave communication systems. Besides, exploiting different beamforming architectures at the mm-Wave base station (BS) generates several directional beams with higher gains to serve the aligned users. By enabling these technological advancements, the next-generation networks may support many services with heterogeneous requirements that lead to having dense users and high data traffic in the cellular network compared to the current trends.

Moreover, driven by the rapid utilization of more wireless terminals and the increment in data traffic, the current wireless networks' energy consumption has been increasing at an alarming rate in recent years. In addition, the upcoming wireless networks' energy consumption can be high due to its support for multiple services and their related applications. Thus, energy-efficient designs are of paramount importance to enable sustainable 5G wireless systems. To this end, the resource allocation scheme targets to improve the EE has become a significant research problem in the design of next-generation wireless networks. In this context, different radio access network (RAN) resource allocation mechanisms have already been proposed in the literature to improve the overall system's EE. Notably, various resource allocation mechanisms aim at enhancing the entire network's throughput while consuming less energy to achieve the quality-of-service (QoS) requirements of users [5], [6]. However, most of the works in the literature focused on orthogonal frequency division multiple access (OFDMA) systems consisting of a single service. Therefore, these works are not directly adaptable to the next-generation networks that support multiple services, including eMBB, mMTC, and URLLC.

Out of the above-mentioned three services, two vital services to be reinforced in the next release of 5G networks are URLLC and eMBB services [7], and hence those are considered in this paper. Accommodation of these services with heterogeneous requirements mandates a dynamic radio resource slicing framework that can ensure the QoS specifications of each service. Towards this achievement, two types of radio resource scheduling approaches have been introduced recently in the 3GPP standards [8] to allocate resource blocks (RBs) for the co-existence of eMBB and URLLC users while satisfying their QoS requirements: (1) puncturing-based scheduling and (2) orthogonal resource scheduling. In the puncturing method, the arrived URLLC traffic is scheduled immediately on the ongoing eMBB transmissions in short transmission time intervals (TTIs). On the contrary, the RBs are pre-reserved for URLLC and eMBB traffic in the orthogonal resource scheduling method. In this work, we consider an orthogonal resource scheduling method for the efficient multiplexing of eMBB and URLLC services on the same radio resources. However, the orthogonal scheduling approach wastes the reserved resources during the absence of URLLC traffic. In order to avoid this resource underutilization problem, the slice-aware resource slicing mechanism proposed in [9] is considered in this work. The RBs of flexible numerologies with different sub-carrier spacing (SCS) have been recently introduced in the 5G communication standards to achieve the QoS requirements of heterogeneous services. Besides, the availability of multi-beam radiation pattern at the mmWave BS in future wireless networks is expected in the next generation of cellular systems. Thus, it is imperative to study the joint beam selection and resource assignment for eMBB and URLLC users in 5G NR. Moreover, the joint beam selection, allocation of RBs with mixed numerologies, and transmit power for the efficient co-existence of eMBB and URLLC data traffic in an energy-efficient way leads to a challenging resource optimization problem. In this paper, we consider the slice-aware resource slicing approach for allocating the beams, RBs of mixed numerologies, and transmit power jointly to the heterogeneous users according to their QoS requirements and packets queue status. We review the related works from the literature and highlight this paper's major contributions in the following sub-sections.

### A. Related works

The RAN resource slicing problems for an efficient multiplexing of URLLC and eMBB services have recently received considerable research attention in the literature. We review some of the relevant works in the following.

A multi-objective resource optimization problem was proposed in [10] to maximize the sum-throughput of eMBB users and weighted sum-throughput of URLLC users under the latency, reliability constraints of URLLC traffic, and the budget power constraints. In [11], the authors proposed a fair scheduling approach for efficient multiplexing of URLLC and eMBB data traffic in 5G downlink, intending to maximize the proportional fairness (PF) for eMBB users while guaranteeing the latency and reliability constraints of URLLC. A heuristic and one-sided matching game algorithm has been studied in [12] for the efficient co-existence of eMBB and URLLC traffic on the shared physical network. Furthermore, to extend the work in [12], the authors introduced a new approach in [13] for multiplexing URLLC and eMBB data traffic on the same radio resources. Precisely, they expressed the co-existence problem as maximization of eMBB's MEAR subject to the scheduling constraints of URLLC traffic. The authors considered an optimization framework in [14] to maximize the data rate of eMBB users while ensuring the reliability specification of URLLC by solving a multi-armed bandit problem. Furthermore, the resource allocation problem was formulated in [15] as an optimization-aided deep reinforcement learning framework to maximize the eMBB data rate subject to the URLLC service's reliability constraint. Moreover, the earlier mentioned works [10]- [15] exploit the puncturing-based mechanism for scheduling the URLLC data traffic on the allocated resources for the eMBB service. The puncturing technique causes the performance reduction of eMBB in terms of reliability and achievable data rate at the arrival of higher URLLC data traffic.

In the other direction, the orthogonal scheduling approach for resource allocation has been considered in the literature to avoid the shortcomings of the puncturing mechanisms. For instance, an orthogonal scheduling mechanism was proposed in [16] for multiplexing of URLLC and eMBB data traffic in an OFDMA-based C-RAN by considering the uncertainty in data traffic load, isolation between slices, and the interference between RRHs. We formulated sum-rate maximization problems in our previous works [17], [18] for sharing the resources orthogonally to eMBB and URLLC services according to their QoS requirements. Furthermore, our other work [19] addressed the problem of joint allocation of RBs and its associated transmit power to the URLLC and eMBB data traffic while satisfying their QoS requirements under the imperfect CSI. However, the above-mentioned works have not formulated the radio resource management (RRM) mechanisms for energy-efficient communications in OFDMA-based mmWave networks consisting of 5G services such as eMBB and URLLC.

On the other hand, some works have studied the RRM techniques in the literature for EE communications in 4G communication scenarios. For instance, a max-min EE optimization problem was studied in [20] for resource allocation in a conventional OFDMA system while ensuring fairness among users, RBs, and transmit power assignment policies. A resource allocation problem was formulated in [21] as an EE maximization problem for optimizing the number of transmitting antennas jointly with the RBs and powers in the downlink OFDMA network. Furthermore, to show the trade-off between network backhaul capacities and EE, a joint sub-carrier and transmit power allocation problem was investigated in [22] for energy-efficient downlink transmissions in OFDMA wireless networks with limited backhaul capacity. The authors of [23] have proposed an optimization problem to allocate radio resources in a MIMO-OFDMA-based LTE system. Specifically, they formulated an EE maximization problem to satisfy the constraints for RB utilization, minimum data-rate constraints for users, and the BS's budget transmission power. However, the works in [20]- [23] have not considered the multiple services. In [24], a resource allocation problem was formulated to maximize the EE performance of an OFDMA-based Het-net while satisfying the constraints for delay-sensitive and delay-tolerant services.

Further, in our other work [25], by considering the mixed-numerology-based time-frequency grid models, we formulated the resource allocation to heterogeneous services as a transmit power minimization problem subject to the QoS constraints of different services. Nevertheless, except [25], the above-mentioned works assumed a fixed numerology-based RBs grid model for scheduling the time-frequency resources to the users, which is suitable to satisfy the stringent QoS requirements of next-generation networks. Furthermore, none of the works considered the joint beam selection, RBs of mixed numerologies, and transmit power allocation in OFDMA-based mmWave DL network with multiple services. In other words, the literature has not addressed the problem of joint beam selection and energy-efficient radio resource allocation in the DL of OFDMA-based mm-Wave network according to URLLC and eMBB traffic arrivals and their QoS requirements. In Table. I, we show how the proposed work in this paper is different from the other works in the literature and our previous work.

### B. Major contributions

Inspired by the above-mentioned observations, we introduce a slice-aware RAN resource-slicing strategy for ensuring the energy-efficient multiplexing of URLLC and eMBB services in the OFDMA-based mmWave DL network. Importantly, we address the EE maximization problem for joint allocation of the beam, RBs of multiple numerologies, and transmit powers to the eMBB and URLLC users according to their data buffer queue status in this paper. In particular, we maximize the overall network EE while considering the constraints for beams and RBs scheduling, power-related, latency-related, and minimum data-rate constraints. Subsequently, we solve the formulated problem using the Dinkelbach-based iterative algorithm, DC programming, and SCA technique. The main contributions of this work are summarized as follows:

- We formulate the joint beam selection and radio resource slicing problem at the BS as an optimization problem to maximize the total network's EE performance while guaranteeing the orthogonality, resources scheduling, power-related constraints, and QoS requirements of eMBB and URLLC users. Specifically, a delay constraint derived from the effective bandwidth method is considered in the problem formulation to ensure the latency requirement of URLLC users in delivering the data packets. Besides, each service's reliability requirement is satisfied using an adaptive MCS scheme tailored according to their Block Error Probability (BLEP).

- The formulated EE maximization problem is a MINLFP problem that is intractable and difficult to solve due to its non-convex and binary nature, which arises from the fractional form of the objective function and the binary decision variables for RBs and beams allocation. We transform the problem into a more tractable form and utilize a two-loop procedure to provide a feasible solution. The main optimization problem (i.e., beam selection and resource allocation) is solved in the inner loop. The outer loop is performed using the Dinkelbach-based iterative method to obtain an improved solution in every iteration until convergence.

- The main optimization problem is decomposed into two sub-problems to provide the low-complexity solution: 1) optimal beam selection and 2) joint RBs and transmit power allocation. After solving the first sub-problem, the obtained optimal beams set is sent to the second sub-problem as an input. Further, the second sub-problem is solved by leveraging the big-M formulation theory, DC programming, and SCA for the given beams set.

- Finally, we compare the performance of the proposed method against baseline schemes through simulation results. In particular, we show the performances of RAN slicing mechanisms with the mixed and fixed-numerology-based RBs grid models in terms of achievable EE, packet latencies, data rates, total sum rate, and computational complexity.

The remainder of the paper is organized as follows. We provide the details of the network model, radio resource grid models, channel model, and link adaptation process for the mmWave DL OFDMA system in Section II. In Section III, we first describe the power consumption model for the considered system model. Next, the EE performance metric is defined and formulated for the slice-aware radio resource allocation problem as an EE maximization problem. Subsequently, the formulated non-convex optimization problem is solved using the Dinkelbach iterative algorithm, DC programming, and SCA in Section IV. Section V presents the simulation results and their related analysis. Finally, conclusions are drawn in Section VI.

## II. SYSTEM MODEL

### A. Network model

We consider a single-cell-based millimeter-wave (mmWave) downlink (DL) network, where a BS has $\mathcal{M}$ fixed predefined beams for transmitting the data to $\mathcal{K}$ eMBB and $\mathcal{L}$ URLLC users, as illustrated in Fig. 1. The sets of eMBB and URLLC users are denoted by $\mathcal{U}_{mbb} = \{1, 2, 3, ...., \mathcal{K}\}$, and $\mathcal{U}_{llc} = \{1, 2, 3, ..., \mathcal{L}\}$, respectively. Also, we assume the OFDMA scheme for the orthogonal allocation of radio RBs to the users. We assume time-frequency RBs with flexible numerologies to satisfy the users' QoS requirements. The complete details of considered frame numerologies and resource grid models are provided in the following sub-sections. The BS is assumed to utilize a sub-connected hybrid beamforming architecture [34], wherein each RF chain is connected to a group of antenna elements, i.e., called sub-array. Each of $\mathcal{M}$ sub-arrays can generate a beam and steer that towards one of the $\mathcal{M}$ pre-defined directions, indexed by $\Theta = \{\theta_1, \theta_2, ..., \theta_{\mathcal{M}}\}$. Furthermore, each user is assumed to be equipped with an omnidirectional antenna that induces a unit gain in all directions. We also assume that each user can estimate the channel quality of radiated beams using the provided channel state information-reference signal (CSI-RS) on every RB and subsequently compute the SINR levels of each beam. The BS serves the users by jointly allocating the beam, RBs, and its associated transmit powers according to their acquired SINR levels on RBs and data traffic arrivals. For each user, all the generated data packets from the higher layers are received at the BS and stored in their corresponding data buffers until they get served, as shown in Fig. 1 (b). Moreover, all the available packets in the BS's data buffer are served based on a first-come-first-serve basis for every individual service.

TABLE I: Comparison of relevant EE works

| Works | Carrier frequency | | RB's Grid model | | Objective function | | Services | | Radio resources | | |
|---|---|---|---|---|---|---|---|---|---|---|---|
| | Sub-6GHz | mmWave | Fix | Mix | Power minimization | EE maximization | E | U | Beams | RBs | Pow |
| [20]- [24] | ✓ | | ✓ | | | ✓ | ✓ | | | ✓ | ✓ |
| Previous work [25] | ✓ | | ✓ | ✓ | ✓ | | ✓ | ✓ | | ✓ | ✓ |
| This work | | ✓ | ✓ | ✓ | | ✓ | ✓ | ✓ | ✓ | ✓ | ✓ |

TABLE II: Summary of notations

| Symbol | Definition |
|---|---|
| $\mathcal{U}_{mbb}$ | Set of available eMBB users |
| $\mathcal{U}_{llc}$ | Set of available URLLC users |
| $\mathcal{K}, \mathcal{L}$ | Total eMBB, URLLC users in the network |
| $\lambda_u$ | URLLC packets arrival rate per user |
| $B_{pkt}$ | URLLC packet size (in bytes) |
| $\mathcal{T}_{RB_i}$ | Time-slot of an RB with numerology '$i$' |
| $\mathcal{T}_i, \mathcal{F}_i$ | No. of TTIs and RBs per TTI in a slice with numerology '$i$' |
| $\mathcal{W}_{RB_i}$ | Bandwidth of an RB with numerology '$i$' |
| $r^{u,\theta}_{t_i,f_i}$ | Data bits of $u^{th}$ user that can be sent on the RB $(t_i, f_i)$ using beam '$\theta$' |
| $\Xi^{u,\theta}_{t_i,f_i}$ | SINR of $u^{th}$ user on the RB $(t_i, f_i)$ of beam '$\theta$' |
| $h^{u,\theta}_{t_i,f_i}$ | Channel between the BS and $u^{th}$ user on the RB $(t_i, f_i)$ for configured beam '$\theta$' |
| $\beta$ | Blockage parameter |
| $P_{max}$ | Available total power per beam at BS |
| $x^u_{t_i,f_i}$ | Binary decision variable |
| $p^u_{t_i,f_i}$ | Allocated power to the RB $(t_i, f_i)$ |
| $\alpha_{los}, \alpha_{nlos}$ | LoS and NLoS Path loss exponents |
| $\mathcal{D}_{BS,u}$ | Distance between the BS and user |
| $\mathcal{N}_o, \mathcal{Z}(\cdot)$ | Noise Power, SE of the selected MCS |
| $Q_u$ | $u^{th}$ URLLC user's queue length |
| $e_u$ | Required data rate for $u^{th}$ eMBB user |
| $T_f, D_u$ | Transmission delay, queuing delay |
| $\beta_u$ | Required data rate for $u^{th}$ URLLC user |
| $\mathcal{B}_E, \mathcal{B}_U$ | BLEP target for eMBB, URLLC services |
| $\phi_u$ | Phase angle of $u^{th}$ user |
| $\lambda_1, \nabla V(\cdot)$ | Penalty parameter, Gradient of $V(\cdot)$ |
| $\mathcal{G}_\theta$ | Antenna radiation pattern of beam '$\theta$' |
| $\mathcal{M}_\theta, m_\theta$ | Major and side lobe gains of beam '$\theta$' |
| $\mathcal{A}_{los}, \mathcal{A}_{nlos}$ | LoS and NLoS Path loss intercepts |

*Queue model:* The users associated with the URLLC service are assumed to generate the packets of $B_{pkt}$ bytes by following the Poisson type traffic model (i.e., FTP 3 model [1] ) with the arrival rate of $\lambda_l$ [packets/sub-frame]. The users demanding the eMBB service are assumed to generate full-buffer traffic with continuous packet arrivals. For eMBB services, it is essential to maintain the queuing system's stability. For URLLC services, we need to satisfy the queuing delay bound, the threshold level of queuing delay bound violation probability, and the transmission delay bound.

### B. 5G Time-frequency Frame Numerology

Unlike the 4G LTE standard, the 5G NR utilizes the flexible numerology that permits SCS to scale $2^\mu \times 15$ kHz [26] mainly (i) to cover the wide range of operating frequencies such as sub-6GHz and mmWave and (ii) for satisfying different QoS requirements of heterogeneous services. According to the 5G NR terminology, the next-generation wireless system supports the summarized numerologies in Table III. As provided in [27], the only SCSs considered for data transmission by accessing frequencies above sub-6 GHz are 60 kHz and 120 kHz, i.e., $2^\mu \times 15$ kHz, with $\mu = \{2, 3\}$, and we will consider only these two SCSs in this work. Since the time duration of the frame and sub-frame (i.e., 10 ms and 1 ms) are preserved as in LTE, by increasing the numerology index, the number of time-slots in every sub-frame increases, as provided in Table III. Moreover, according to the LTE standard, the minimum time-frequency scheduling unit in the OFDMA-based resource allocation is an RB which consists of 12 sub-carriers and 14 OFDM symbols in the time-duration of $\mathcal{T}_{RB}$ ms. With $\mu = \{2, 3\}$, each RB's bandwidth and the duration of time slot or transmission time interval (TTI) are given by

$$\mathcal{W}_{RB} = 2^\mu \times 12 \times 15 \text{ kHz}, \text{ and } \mathcal{T}_{RB} = \frac{1}{2^\mu} \text{ ms}, \ \mu = \{2, 3\}; \quad (1)$$

Contrary to the LTE standard, in 5G NR, each mini-slot may consist of 7, 4, or 2 OFDM symbols and start the transmissions immediately without waiting for slot boundaries to achieve low latency in data packets delivering. In this work, as in our previous work [25], we consider (a) fixed numerology and (b) mixed-numerology-based time-frequency resource grid models for allocating the resources to the users requesting different services. The fixed-numerology-based grid model has RBs of numerology with the same SCS (i.e., either 60 kHz or 120 kHz) as shown in Fig. 2 (a). In contrast, the mixed-numerology-based grid has multiple bandwidth parts (BWPs), where each BWP consists RBs with a specific numerology (i.e., BWP 1 with 60 kHz and BWP 2 with 120 kHz), as shown in Fig. 2 (b). Further, the neighboring RBs with different numerologies are separated with a guard band to avoid inter-numerology interference (INI).

### C. Antenna array gain, Channel Model and SINR Computation

*Antenna Array Gain*: Motivated by the approximation model provided in [29], [4] for antenna array gains, we also approximate each sub antenna array gain at the BS utilizing only two constant parameters, i.e., major-lobe gain $\mathcal{M}_\theta$ and minor-lobe gain $m_\theta$. The received gain for the associated and interfering links is determined using the major and minor lobe gain parameters that are specified by the 3GPP antenna radiation pattern, as given in the following

$$\mathcal{G}_\theta(\phi_u) = \mathcal{M}_\theta, \text{and } \mathcal{G}_{\theta'}(\phi_u) = \begin{cases} \mathcal{M}_{\theta'} & \text{w.p. } \dfrac{\Phi_{\text{Tx}}}{2\pi} \\ m_{\theta'} & \text{w.p. } \left(1 - \dfrac{\Phi_{\text{Tx}}}{2\pi}\right), \end{cases}$$

where the major-lobe and side-lobe gains are computed as $\mathcal{M}_\theta = 10^{0.8} n_{Tx}$, $m_\theta = 1/\sin^2\left(\frac{3\pi}{2\sqrt{n_{Tx}}}\right)$, $n_{Tx}$ represents the number of antennas in each sub-antenna array at the BS, $\Phi_{Tx}$ is the half-power beam-width of the major lobe, and $\phi_u$ represents the phase angle of the user 'u'. Note that we assume the user has a single omnidirectional antenna, which induces a unit gain in all directions.

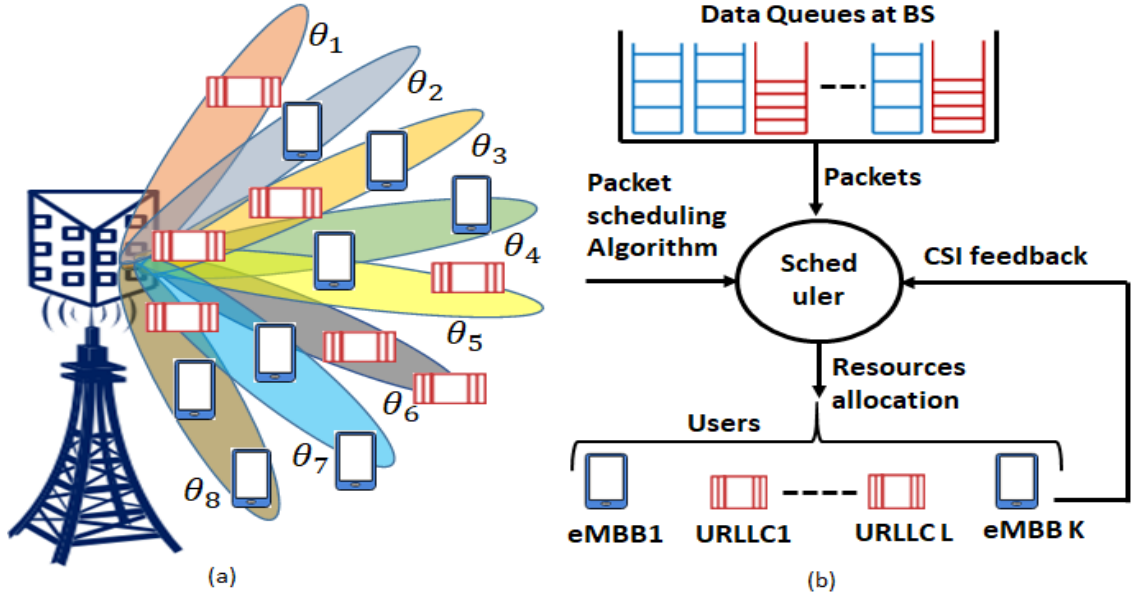

Fig. 1: Illustration of (a) the single-cell network serving eMBB and URLLC users with aligned beams, and (b) generalized scheduling model for DL communications.

TABLE III: Flexible Numerologies for 5G NR

| Index ($\mu$) | SCS (in kHz) | Time-slot (in ms) | RB's bandwidth (in kHz) |
|---|---|---|---|
| 0 | 15 | 1 | 180 |
| 1 | 30 | 0.5 | 360 |
| 2 | 60 | 0.25 | 720 |
| 3 | 120 | 0.125 | 1440 |
| 4 | 240 | 0.06125 | 2880 |

*Channel model*: We assume that channels in the configured beams are following the block fading in both frequency and time domains. The channel coefficients on different RBs allocated to users are assumed to be independent and identically distributed. Accordingly, for analytical tractability, like the work in [28], we adopt the widely used extended Saleh-Valenzuela channel model for mmWave transmission links.

Since the mmWave network experiences signal strength degradation due to the effect of severe blockages and path loss, we consider the following path loss model for LoS and NLoS communication links

$$\mathcal{L}(\mathcal{D}) = \begin{cases} \mathcal{A}_{los}\mathcal{D}^{-\alpha_{los}} & \text{w.p. } P_{los}(\mathcal{D}) \\ \mathcal{A}_{nlos}\mathcal{D}^{-\alpha_{nlos}} & \text{w.p. } P_{nlos}(\mathcal{D}), \end{cases} \quad (3)$$

where $\mathcal{A}_{los}$, $\mathcal{A}_{los}$, $\alpha_{los}$, and $\alpha_{nlos}$ represent the path loss intercepts at the unit distance and the path loss exponents for LoS and NLoS links, and for a provided link distance $\mathcal{D}$, the probabilities of LoS and NLoS states occurrence for each channel can be provided as a function of $\mathcal{D}$ as

$$P_{los}(\mathcal{D}) = e^{-\beta\mathcal{D}}, \text{and } P_{nlos}(\mathcal{D}) = 1 - P_{los}(\mathcal{D}), \quad (4)$$

where $\beta$ is a blocking parameter (i.e., estimated by the average sizes and density of blockages).

*SINR computation*: By letting $h_{t_i,f_i}^{u,\theta}$ is a channel coefficient on the RB scheduled to the $u$-th user for the configured beam $\theta$, $p_{t_i,f_i}^u$ is the assigned transmit power to each RB, and using the earlier mentioned notations and assumptions, the received signal-to-interference plus noise ratio (SINR) of the $u$-th user on the scheduled RB $(t_i, f_i)$ for the configured beam $\theta$ can be expressed as

$$\Xi_{t_i,f_i}^{u,\theta} = \frac{p_{t_i,f_i}^u |h_{t_i,f_i}^{u,\theta}|^2 \mathcal{G}_\theta(\phi_u)\mathcal{L}(\mathcal{D}_{BS,u})}{\mathcal{N}_o + \gamma_{int}}, \quad (5)$$

where $\mathcal{D}_{BS,u}$, $\gamma_{int} = \displaystyle\sum_{\theta' \neq \theta, u' \neq u} |h_{t_i,f_i}^{u',\theta'}|^2 \mathcal{G}_{\theta'}(\phi_u) p_{t_i,f_i}^{u',\theta'} \mathcal{L}(\mathcal{D}_{BS,u'})$ and $\mathcal{N}_o$ represent the distance between the associated BS and $u$-th user, the interference term that is occurred when the user receives a significant power from the side lobes of

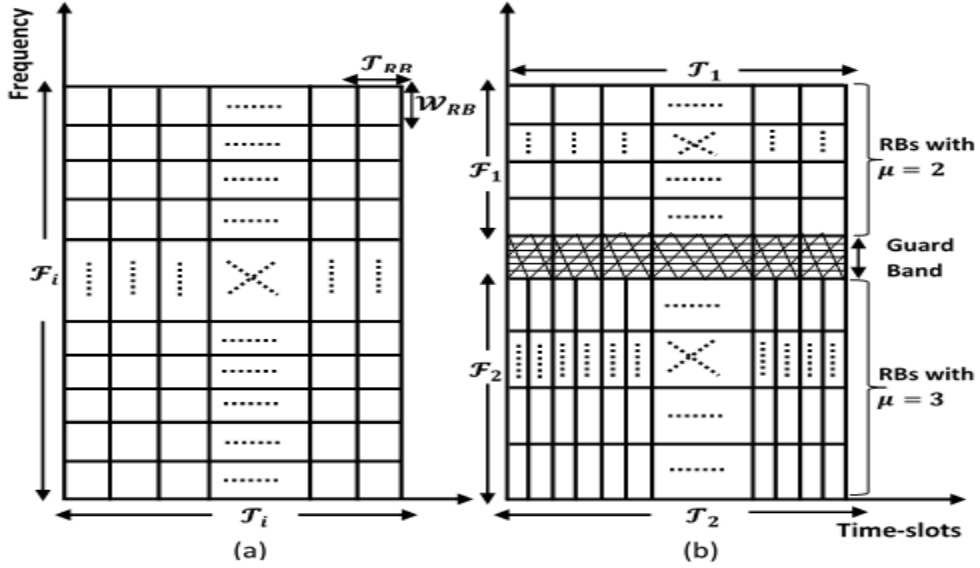

Fig. 2: Time-frequency resource blocks with (a) fixed numerology and (b) mixed numerology.

other beams, and the power of additive white Gaussian noise (AWGN) at users, respectively.

### D. Link adaptation

The link adaptation and packet scheduling functionalities play a vital role in achieving the QoS requirements of users. This dynamic link adaptation process needs to be invoked for every user regardless of the requesting service type by adjusting the utilized MCS per DL transmission. The main steps involved in this link adaptation process are channel quality indicator (CQI) measuring and reporting. The BS configures the UE through the RRC signaling with one or many CSI reporting and resource settings as a first step. Afterward, the UE executes the CQ measurements using the specified CSI-RS on every RB. Next, every measurement is filtered and utilized to compute the user's achievable SINR on all the available RBs. According to the user's received SINR on the RB and BLER threshold of its requesting service, the BS schedules the RB to a user by mapping the MCS from Table II given in [25]. Note that the measured CQI and other reports, such as the precoding matrix indicator and rank indicator, are sent to the transmitter through the CSI feedback.

## III. Problem Formulation

By using the earlier mentioned notations and assumptions, the achievable transmission data rate of the user '$u$' on the scheduled RB $(t_i, f_i)$ for the configured beam '$\theta$' is given by

$$\tilde{r}^{u,\theta}_{t_i,f_i} = \mathcal{W}_{RB_i} \cdot \mathcal{T}_{RB_i} \cdot \mathcal{Z}(\cdot) \quad \text{[bits/RB]} \tag{6}$$

where $\mathcal{Z}(\cdot)$ represents the spectral efficiency (SE) of the received MCS from Table according to the estimated SINR level. The URLLC service requires to maintain highly reliable communication links compared to the eMBB service. Thus, in this paper, two separate MCSs are considered, and the corresponding SINR levels of MCS for different BLEP targets are provided in Table [17], [31]. In particular, BLEPs

$\mathcal{B}_U = 10^{-5}$, and $\mathcal{B}_E = 10^{-3}$ are considered for URLLC and eMBB, respectively. However, the presence of step-wise function $\mathcal{Z}(\cdot)$ in $\tilde{r}^{u,\theta}_{t_i,f_i}$ causes the mathematical intractability and induces complexity in solving of the optimization problem when it exists in the objective function or constraints. Towards addressing this problem, motivated by [17], utilizing the target BLER and received SINR, we consider the following approximations for URLLC and eMBB services that can be expressed as

$$\mathcal{Z}_L(\Xi^{u,\theta}_{t_i,f_i}, \mathcal{B}_U) = \log_2\left(1 + \frac{\Xi^{u,\theta}_{t_i,f_i}}{\Gamma_U}\right), \text{and}$$

$$\mathcal{Z}_E(\Xi^{u,\theta}_{t_i,f_i}, \mathcal{B}_E) = \log_2\left(1 + \frac{\Xi^{u,\theta}_{t_i,f_i}}{\Gamma_E}\right), \tag{7}$$

where $\Gamma_E = \frac{-\ln(5\mathcal{B}_E)}{1.5}$ and $\Gamma_U = \frac{-\ln(5\mathcal{B}_U)}{0.45}$ provide the values of the SINR gaps, respectively. Now, by using the SE approximation functions, the bit rate of each RB can be re-written as

$$r^{u,\theta}_{t_i,f_i} = \mathcal{W}_{RB_i} \cdot \mathcal{T}_{RB_i} \cdot \mathcal{Z}_s(\cdot) \quad \text{[bits/RB]} \tag{8}$$

By using (8), the achievable data rate of the $u$-th user is computed as

$$R_u = \sum_{t_i=1}^{N_i} \sum_{f_i=1}^{\mathcal{F}_i} \sum_{\theta=1}^{\mathcal{M}} x^u_{t_i,f_i} \ I_{\theta,u} \ r^{u,\theta}_{t_i,f_i} \tag{9}$$

where $x^u_{t_i,f_i}$ is a binary variable, which is 1 if the user '$u$' served by the beam '$\theta$' is scheduled on the RB $(t_i, f_i)$, and otherwise it is 0, $I_{\theta,u}$ is an another binary decision variable, which is 1 if the beam '$\theta$' is configured to serve the user '$u$', and otherwise, it is 0. Mathematically, the earlier mentioned two binary variables are expressed as

$$x^u_{t_i,f_i} = \left\{ \begin{array}{ll} 1; & \text{If RB is allocated to the user '}u\text{'} \\ 0; & \text{Otherwise} \end{array} \right\},$$

$$I_{\theta,u} = \left\{ \begin{array}{ll} 1; & \text{If the user '}u\text{' is served by a beam '}\theta\text{'} \\ 0; & \text{Otherwise} \end{array} \right\} \tag{10}$$

Now, the achievable data rate of the overall network (i.e., accumulation of achievable data rates of all active users) is written as

$$\mathcal{R}(\mathcal{X}, \mathcal{P}, \mathcal{I}) = \sum_{u=1}^{K+L} R_u \qquad (11)$$

where $\mathcal{X}, \mathcal{I},$ and $\mathcal{P}$ are the RB, beam, and transmit power assignment policies, respectively.

*QoS requirement for eMBB service:* We need to assure the queue stability for eMBB services, which is achieved by maintaining the achievable data rate equal to or higher than the data arrival rate of every eMBB user, i.e.,

$$R_u \geq e_u, \forall u \in \mathcal{U}_{mbb}. \qquad (12)$$

where $e_u$ represents the data rate requirement of $u$-th eMBB user.

*QoS requirement for URLLC service:* A delay bound and a threshold of delay bound violation probability need to be satisfied for URLLC services. To ensure these requirements mentioned earlier for URLLC services, the effective BW concept is utilized in [32]. In order to achieve the queuing delay bound of the $u$-th URLLC user, the achievable data rate should not be lower than the effective BW of the user's arrival process. The effective BW of the Poisson process is expressed as [33],

$$E_u^B(\psi_u) = \frac{\lambda_u}{T_f \psi_u}(e^{\psi_u} - 1)\text{(packets/s)}, \qquad (13)$$

where $\psi_u$ is the QoS exponent, which is expressed as

$$\psi_u = \ln\left[\frac{T_f \ln(1/\epsilon_u)}{\lambda_u D_u} + 1\right] \qquad (14)$$

In other words, to ensure the queuing delay bound and the probability of queuing delay bound violation, the following constraint should be satisfied [33],

$$R_u \geq \tau_u, u \in \mathcal{U}_{llc}. \qquad (15)$$

where $\tau_u = \min(Q_u, T_f E_u^B(\psi_u))B_{pkt}$, $Q_u$ is the queue length (i.e., in packets) of the $u$-th URLLC user, $T_f$ is the transmission frame duration, and $B_{pkt}$ is the number of bits per URLLC packet.

### A. Power-consumption model

In order to model an energy-efficient resource allocation mechanism, it is crucial to include the total network's power consumption in the objective function of the optimization problem. Inspired by works [34], [35], we model the total power consumption of the transmission as the accumulation of dynamic and one static term:

$$\mathcal{PC}(\mathcal{P}, \mathcal{I}, \mathcal{X}) = \frac{1}{\zeta} \sum_{i \in \{1,2\}} \sum_{u=1}^{K+L} \sum_{t_i=1}^{N} \sum_{f_i=1}^{\mathcal{F}_i} \sum_{\theta=1}^{\mathcal{M}} x^u_{t_i,f_i} I_{\theta,u} \ p^u_{t_i,f_i} +$$

$$n_{Tx}P_c \sum_{i \in \{1,2\}} \sum_{u=1}^{K+L} \sum_{t_i=1}^{N} \sum_{f_i=1}^{\mathcal{F}_i} \sum_{\theta=1}^{\mathcal{M}} x^u_{t_i,f_i} I_{\theta,u} + P_s \qquad (16)$$

where $\zeta \in (0,1)$ represents the power amplifier's drain efficiency, $P_c$ and $P_s$ represent each antenna's power consumption for signal processing on each RB, and the static circuit power consumption, respectively.

### B. EE maximization Problem Formulation

In this section, we formulate an optimization problem to maximize the total network EE by considering the joint beam, RB, and power allocation to the active eMBB and URLLC users in the network. EE is a metric to measure how efficiently the system consumes energy. Thus, the wireless communication systems' EE is formulated as a ratio of the total achievable rate to the overall network's power consumption. The EE of the considered wireless system is written as

$$\eta_{EE}(\mathcal{X}, \mathcal{P}, \mathcal{I}) = \frac{\mathcal{R}(\mathcal{X}, \mathcal{P}, \mathcal{I})}{\mathcal{PC}(\mathcal{X}, \mathcal{P}, \mathcal{I})} \qquad (17)$$

To obtain optimal beam, RB, and transmit power allocation policies while maximizing the earlier mentioned total network EE subject to the QoS requirements of all the users, scheduling, and power-related constraints, we formulate the following optimization problem

$$\mathcal{OP}_1 : \max_{\{\mathcal{X}, \mathcal{P}, \mathcal{I}\}} \eta_{EE}(\mathcal{X}, \mathcal{P}, \mathcal{I}) \qquad (18)$$

subject to
$$\mathcal{C}_1 : I_{\theta,u} \in \{0,1\}, \forall \theta, u;$$
$$\mathcal{C}_2 : x^u_{t_i,f_i} \in \{0,1\}, \forall u, t_i, f_i;$$
$$\mathcal{C}_3 : \sum_{\theta \in \Theta} I_{\theta,u} \leq 1, \forall u;$$
$$\mathcal{C}_4 : \sum_{u \in \mathcal{U}} x^u_{t_i,f_i} I_{\theta,u} \leq 1, \forall t_i, f_i, \theta;$$
$$\mathcal{C}_5 : \sum_{\theta \in \Theta} \sum_{t_1=1}^{N_1} \sum_{f_1=1}^{\mathcal{F}_1} I_{\theta,u} x^u_{t_1,f_1} r^{u,\theta}_{t_1,f_1} \geq e_u, \forall u \in U_{mbb};$$
$$\mathcal{C}_6 : \sum_{\theta \in \Theta} \sum_{t_2=1}^{N_2} \sum_{f_2=1}^{\mathcal{F}_2} I_{\theta,u} x^u_{t_2,f_2} r^{u,\theta}_{t_2,f_2} \geq \tau_u, \forall u \in U_{llc};$$
$$\mathcal{C}_7 : \sum_{\theta \in \Theta} \sum_{t_2=1}^{N_2} \sum_{f_2=1}^{\mathcal{F}_2} I_{\theta,u} x^u_{t_2,f_2} r^{u,\theta}_{t_2,f_2} \geq \zeta_u, \forall u \in U_{mbb};$$
$$\mathcal{C}_8 : p^u_{t_i,f_i} \geq 0, \forall u, t_i, f_i;$$
$$\mathcal{C}_9 : \sum_{i \in \{1,2\}} \sum_{t_i=1}^{N_i} \sum_{f_i=1}^{\mathcal{F}_i} \sum_{u=1}^{K+L} I_{\theta,u} x^u_{t_i,f_i} p^u_{t_i,f_i} \leq P_{max}, \forall \theta;$$

The constraints are explained in the following

- The beam selection constraint $\mathcal{C}_1$ ensures that if the beam '$\theta$' is configured in the time-frame to serve the user '$u$', the value of the variable $I_{\theta,u}$ is 1, and otherwise, it is 0.
- The RBs scheduling constraint $\mathcal{C}_2$ specifies that the value of the variable $x^u_{t_i,f_i}$ is 1 when the user '$u$' is scheduled on the RB $(t_i, f_i)$, and otherwise, it is 0.
- Constraint $\mathcal{C}_3$ confirms that every user associates only to the single beam to avoid the decoding complexity.
- The orthogonality constraint $\mathcal{C}_4$ ensures to schedule each RB to the single user serving by the configured beam '$\theta$'.
- Constraint $\mathcal{C}_5$ guarantees the minimum required data rate for every scheduled eMBB user.
- Constraint $\mathcal{C}_6$ satisfies the requirements of queuing delay bound and delay bound violation probability for every scheduled URLLC user.
- Constraint $\mathcal{C}_7$ ensures that the eMBB users may schedule on the underutilized RBs of URLLC slice to enhance their data rates.

- Constraint $\mathcal{C}_8$ enforces that the assigned transmit power to the scheduled user 'u' on every RB should be positive.
- Constraint $\mathcal{C}_9$ specifies that the accumulation of the assigned transmission powers to all the available RBs in a time-frame serving by the beam '$\theta$' should be lesser or equal to the maximum transmit power available to the beam '$\theta$' at the BS.

*Remarks:*

- For some parameter settings, the optimization problem $\mathcal{OP}1$ may not have any feasible solution set due to conflicting constraints. In such cases, parameters in constraints need to adjust (i.e., admission control) to make the problem feasible. Without loss of generality, we consider that all the constraints in the optimization problem $\mathcal{OP}1$ can be satisfied to assure a feasible solution.

The optimization problem $\mathcal{OP}1$ is a mixed-integer non-linear fractional programming problem (MINLFP) due to the involvement of binary constraints $\mathcal{C}_1$, $\mathcal{C}_2$, rate constraints $\mathcal{C}_5 - \mathcal{C}_7$, and the fractional form of the objective function. This problem is difficult to solve directly. Therefore, we propose an iterative algorithm based on the Dinkelbach method to solve the earlier mentioned optimization problem $\mathcal{OP}1$ in the following section.

## IV. PROPOSED SOLUTION

In this section, first, we transform the MINLFP problem into the MINLP problem. Subsequently, we provide a two-loop algorithm to solve the transformed optimization problem globally. The Dinkelbach method is used in the outer loop to estimate the optimal $q$ value iteratively and in the inner loop, the original problem is solved by decomposing it into two sub-problems.

As a first step for solving $\mathcal{OP}_1$, we exploit the following Theorem to transform MINLFP into MINLP as

**Theorem 1:** The maximum EE $q^* = \frac{\mathcal{R}(\mathcal{X}, \mathcal{P}, \mathcal{I})}{\mathcal{PC}(\mathcal{X}, \mathcal{P}, \mathcal{I})}$ is obtained if and only if

$$\max_{\{\mathcal{X}, \mathcal{P}\}} \ \mathcal{R}(\mathcal{X}, \mathcal{P}, \mathcal{I}) - q^* \mathcal{PC}(\mathcal{X}, \mathcal{P}, \mathcal{I})$$
$$= \mathcal{R}(\mathcal{X}^*, \mathcal{P}^*, \mathcal{I}^*) - q^* \mathcal{PC}(\mathcal{X}^*, \mathcal{P}^*, \mathcal{I}^*) = 0, \quad (19)$$

for $\mathcal{R}(\mathcal{X}, \mathcal{P}, \mathcal{I}) \geq 0$ and $\mathcal{PC}(\mathcal{X}, \mathcal{P}, \mathcal{I}) > 0$.

*Proof:* The proof follows a similar approach as in [36]. ∎

Theorem 1 indicates that for a fractional optimization problem, an equivalent optimization problem exists with an objective function in the subtractive form. Then, the optimization problem $\mathcal{OP}1$ can be reformulated using Theorem 1 as

$$\mathcal{OP}2 : \max_{\{\mathcal{X}, \mathcal{P}, \mathcal{I}\}} \ \mathcal{R}(\mathcal{X}, \mathcal{P}, \mathcal{I}) - q \mathcal{PC}(\mathcal{X}, \mathcal{P}, \mathcal{I}) \quad (20)$$
$$\text{subject to } \ \mathcal{C}_1 - \mathcal{C}_9$$

The Dinkelbach method is utilized to determine increasing values of $q$ iteratively, determined using the optimal policies received at each iteration via solving the optimization problem $\mathcal{OP}2$, as provided in Algorithm 1. This procedure generates an increasing order of $q$ values, which converges to the optimal value at a super-linear convergence rate. In every $j$-th iteration, the set of optimal policies $(\mathcal{X}^j, \mathcal{P}^j, \mathcal{I}^j)$ is achieved for the provided value of $q_j$ (i.e., obtained using $(\mathcal{X}^{j-1}, \mathcal{P}^{j-1}, \mathcal{I}^{j-1})$).

However, the main step in Algorithm 1 is solving the involved MINLP problem for the provided $q_j$ to get the optimal $(\mathcal{X}^*, \mathcal{P}^*, \mathcal{I}^*)$ in every iteration. In the following section, we provide the procedure for how to solve $\mathcal{OP}_2$ for the given $q_j$.

In order to reduce the computational complexity of the MINLP problem, the original optimization problem is decomposed into two sub-problems: 1) optimal beam selection and 2) joint RB and transmit power assignment. The following iterative algorithm is employed to compute the locally optimal RB, transmit power, and beam allocation.

---

**Algorithm 1** Joint allocation of the beam, RB, and transmission power using Dinkelbach method

---

1: **Initialization:** Iterations: $T_{max}$, tolerance $\sigma > 0$, and Initial points : $\mathcal{X}^{(0)}, \mathcal{P}^{(0)}, \mathcal{I}^{int}$.
2: **Set** iteration: $j = 0$, and $q_j = 0$
3: **Repeat**
4:     Solve the optimization problems in (21) and $\mathcal{OP}4$ to obtain $\{\mathcal{X}^j, \mathcal{P}^j, \mathcal{I}\}$
5:     $q_{j+1} = \frac{\mathcal{R}(\mathcal{X}^j, \mathcal{P}^j, \mathcal{I})}{\mathcal{P}(\mathcal{X}^j, \mathcal{P}^j, \mathcal{I})}$
6:     **Set** $j = j + 1$
7:     **until** $|\mathcal{Y}(q_j)| \leq \sigma$ or $j = T_{max}$
8: **Return** $(\mathcal{X}^*, \mathcal{P}^*, \mathcal{I}^*) = (\mathcal{X}^j, \mathcal{P}^j, \mathcal{I})$

---

The proposed EE maximization problem for the joint allocation of RB, beam, and transmit power to the users is solved using the two loops. In the outer loop, we first transform the fractional form objective function into a subtractive form, i.e., $\mathcal{R}(\mathcal{X}, \mathcal{P}, \mathcal{I}) - q_j \mathcal{PC}(\mathcal{X}, \mathcal{P}, \mathcal{I})$. Subsequently, updates the $q_j$ value in an iterative way by taking the values from the lower block (i.e., called an inner loop). In the inner loop, for the given $q_j$, the major problem is solved by decomposing it into two sub-problems. The solution to each sub-problem has been described in Sections IV-A and IV-B. After getting the optimal $\mathcal{X}^*$, $\mathcal{I}^*$ and $\mathcal{P}^*$ from the sub-problems, those values are fed back to the upper block to update the $q_j$ value as shown in Fig. 3. Many inner and outer iterations are performed to achieve the optimal $(\mathcal{X}^*, \mathcal{I}^*, \mathcal{P}^*)$, and $q_j$ value. For clarity, the complete solution methodology is briefed in Fig. 3.

### A. Beams Selection

In this section, we investigate the beam selection for each user from the available beams at the BS. By assuming RBs and their associated transmit power assignment are attained from the previous iteration, a suitable beam for each user can be identified as

$$I_{\theta, u} = \begin{Bmatrix} 1; & \text{If } \theta = \arg \max_{\theta \in \Theta} R_u(\mathcal{X}^n, \mathcal{P}^n, \phi_u) \\ 0; & \text{Otherwise} \end{Bmatrix} \quad (21)$$

To solve the problem in (21), we need to have the knowledge of achievable SINRs of users on every scheduled RB. Furthermore, we assume that the user feeds back the CSI to the BS after estimating it using the received reference signal from the BS. Note that the CSI includes information about the direction of the highest received signal strength and channel gains.

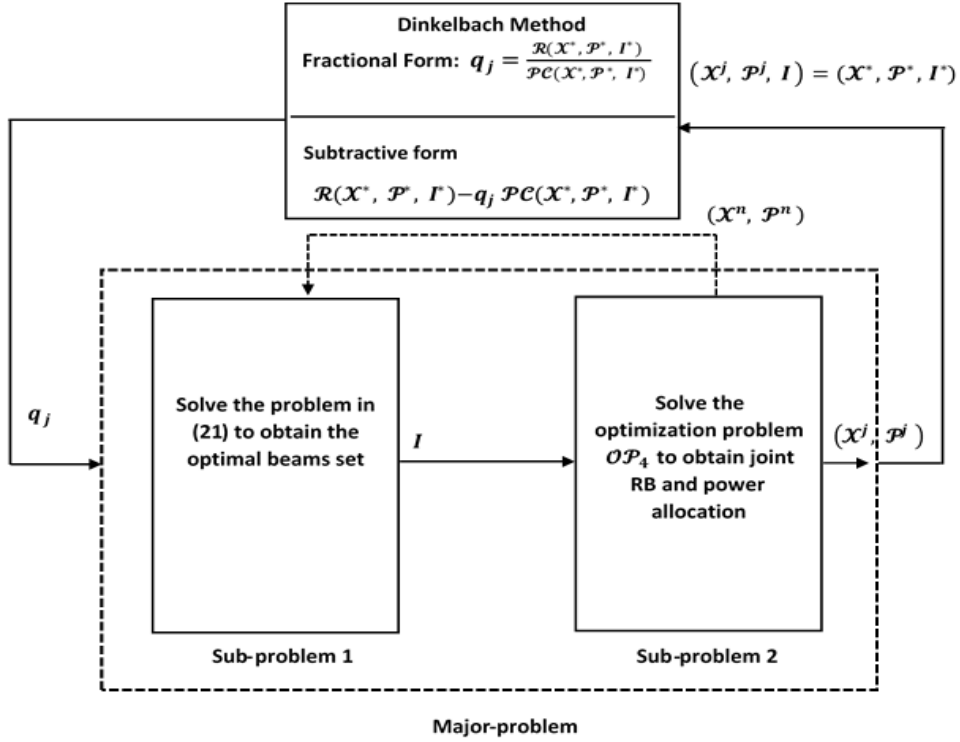

Fig. 3: Block diagram of the summary of solution methodology.

### B. Joint Transmit Power and RB assignment

Assuming the selected beams set from the previous iteration, a solution for the joint RB and transmit power allocation is provided in this section. Though we assume the selected beams set from the previous iteration, the optimization problem $\mathcal{OP}_2$ is still a non-convex problem due to the binary nature of the RBs assignment variable, multiplicative term $x_{t_i,f_i}^u p_{t_i,f_i}^u$ in the objective function and rate constraints. To avoid the binary nature of the RBs assignment variable, we relax the variable between $0$ and $1$ as

$$\dot{\mathcal{C}}_2 : 0 \leq x_{t_i,f_i}^u \leq 1, \forall t_i, f_i, u;$$

In addition, we introduce the following constraint to ensure binary values as

$$\bar{\mathcal{C}}_2 : \sum_{i \in \{1,2\}} \sum_{t_i=1}^{N_i} \sum_{f_i=1}^{\mathcal{F}_i} \sum_{u=1}^{K+L} (x_{t_i,f_i}^u - (x_{t_i,f_i}^u)^2) \leq 0.$$

Further, we leverage the big-M formulation theory to decouple the multiplicative terms in the constraints and objective function as

$$0 < p_{t_i,f_i}^u \leq x_{t_i,f_i}^u \cdot P_{max}, \forall t_i, f_i, u;$$

By using the aforementioned techniques, the problem $\mathcal{OP}_2$ can be modified for joint RB and transmit power allocation as

$$\mathcal{OP}_3 : \max_{\{\mathcal{X},\mathcal{P}\}} \mathcal{R}'(\mathcal{X},\mathcal{P},\mathcal{I}) - q\mathcal{PC}'(\mathcal{X},\mathcal{P},\mathcal{I}) \tag{22}$$

subject to $\dot{\mathcal{C}}_2$, $\bar{\mathcal{C}}_2$, $\mathcal{C}_3$, $\mathcal{C}_4$,

$$\mathcal{C}_5 : \sum_{\theta \in \Theta} \sum_{t_1=1}^{N_1} \sum_{f_1=1}^{\mathcal{F}_1} I_{\theta,u} r_{t_1,f_1}^{u,\theta} \geq e_u, \forall u \in U_{mbb};$$

$$\mathcal{C}_6 : \sum_{\theta \in \Theta} \sum_{t_2=1}^{N_2} \sum_{f_2=1}^{\mathcal{F}_2} I_{\theta,u} r_{t_2,f_2}^{u,\theta} \geq \tau_u, \forall u \in U_{llc};$$

$$\mathcal{C}_7 : \sum_{\theta \in \Theta} \sum_{t_2=1}^{N_2} \sum_{f_2=1}^{\mathcal{F}_2} I_{\theta,u} r_{t_2,f_2}^{u,\theta} \geq \zeta_u, \forall u \in U_{mbb};$$

$$\mathcal{C}_8 : 0 < p_{t_i,f_i}^u \leq x_{t_i,f_i}^u \cdot P_{max}, \forall u, t_i, f_i;$$

$$\mathcal{C}_9 : \sum_{i \in \{1,2\}} \sum_{t_i=1}^{N_i} \sum_{f_i=1}^{\mathcal{F}_i} \sum_{u=1}^{K+L} I_{\theta,u} p_{t_i,f_i}^u \leq P_{max}, \forall \theta;$$

where

$$\mathcal{PC}'(\mathcal{X},\mathcal{P}) = \frac{1}{\zeta} \sum_{i \in \{1,2\}} \sum_{u=1}^{K+L} \sum_{t_i=1}^{N} \sum_{f_i=1}^{\mathcal{F}_i} \sum_{\theta=1}^{\mathcal{M}} I_{\theta,u} \ p_{t_i,f_i}^u +$$

$$n_{Tx} P_c \sum_{i \in \{1,2\}} \sum_{u=1}^{K+L} \sum_{t_i=1}^{N} \sum_{f_i=1}^{\mathcal{F}_i} \sum_{\theta=1}^{\mathcal{M}} x_{t_i,f_i}^u I_{\theta,u} + P_s, \text{and}$$

$$\mathcal{R}'(\mathcal{X},\mathcal{P}) = \sum_{t_i=1}^{N_i} \sum_{f_i=1}^{\mathcal{F}_i} \sum_{u=1}^{K+L} \sum_{\theta=1}^{K+L} I_{\theta,u} \ r_{t_i,f_i}^{u,\theta}.$$

However, due to the presence of interference term in the approximated rate functions, the optimization problem $\mathcal{OP}_3$ is still non-convex. To obtain concave approximations for the constraints $\mathcal{C}_5 - \mathcal{C}_7$, we use the majorization minimization (MM) procedure [37] and construct a surrogate function for $\mathcal{Q}(\mathcal{X},\mathcal{P},\mathcal{I})$ using the first-order Taylor approximation as

$$\tilde{\mathcal{C}}_5 : \sum_{\theta \in \Theta} \sum_{t_1=1}^{N_1} \sum_{f_1=1}^{\mathcal{F}_1} \mathcal{W}_E(\mathcal{X}, \mathcal{P}, \mathcal{I})$$
$$- \sum_{\theta \in \Theta} \sum_{t_1=1}^{N_1} \sum_{f_1=1}^{\mathcal{F}_1} \tilde{\mathcal{Q}}_E(\mathcal{X}, \mathcal{P}, \mathcal{I}) \geq e_u, \forall u \in U_{mbb}. \quad (23)$$

$$\tilde{\mathcal{C}}_6 : \sum_{\theta \in \Theta} \sum_{t_2=1}^{N_2} \sum_{f_2=1}^{\mathcal{F}_2} \mathcal{W}_L(\mathcal{X}, \mathcal{P}, \mathcal{I})$$
$$- \sum_{\theta \in \Theta} \sum_{t_2=1}^{N_2} \sum_{f_2=1}^{\mathcal{F}_2} \tilde{\mathcal{Q}}_L(\mathcal{X}, \mathcal{P}, \mathcal{I}) \geq \gamma_u, \forall u \in U_{llc}. \quad (24)$$

$$\tilde{\mathcal{C}}_7 : \sum_{\theta \in \Theta} \sum_{t_2=1}^{N_2} \sum_{f_2=1}^{\mathcal{F}_2} \mathcal{W}_E(\mathcal{X}, \mathcal{P}, \mathcal{I})$$
$$- \sum_{\theta \in \Theta} \sum_{t_2=1}^{N_2} \sum_{f_1=1}^{\mathcal{F}_2} \tilde{\mathcal{Q}}_E(\mathcal{X}, \mathcal{P}, \mathcal{I}) \geq \zeta_u, \forall u \in U_{mbb}. \quad (25)$$

where

$$\mathcal{W}_s(\mathcal{X}, \mathcal{P}, \mathcal{I}) = I_{\theta, u} \log_2 \left( p_{t_i, f_i}^u |h_{t_i, f_i}^{u, \theta}|^2 \mathcal{G}_\theta(\phi_u) \mathcal{L}(\mathcal{D}_{BS, u}) + \right.$$
$$\left. \left( \sum_{\theta' \neq \theta} |h_{t_i, f_i}^{u, \theta'}|^2 \mathcal{G}_{\theta'}(\phi_u) \, p_{t_i, f_i}^u \mathcal{L}(\mathcal{D}_{BS, u}) + \mathcal{N}_o \right) \Gamma_s \right),$$

$$\tilde{\mathcal{Q}}_s(\mathcal{X}, \mathcal{P}, \mathcal{I}) = \mathcal{Q}(\mathcal{X}^{(j-1)}, \mathcal{P}^{(j-1)}, \mathcal{I}^{(j-1)}) +$$
$$\nabla_P \mathcal{Q}^T(\mathcal{X}^{(j-1)}, \mathcal{P}^{(j-1)}, \mathcal{I}^{(j-1)})(\mathcal{P} - \mathcal{P}^{(j-1)}), \text{and}$$

$$\mathcal{Q}_s(\mathcal{X}, \mathcal{P}, \mathcal{I}) = I_{\theta, u} \log_2 \left( \left( \sum_{\theta' \neq \theta} |h_{t_i, f_i}^{u, \theta'}|^2 \mathcal{G}_{\theta'}(\phi_u) \, p_{t_i, f_i}^u \right. \right.$$
$$\left. \left. \mathcal{L}(\mathcal{D}_{BS, u}) + \mathcal{N}_o \right) \Gamma_s \right), s \in \{E, L\}. \quad (26)$$

Similarly, we can re-write the rate term of the objective function of $\mathcal{OP}2$ as

$$\mathcal{R}''(\mathcal{X}, \mathcal{P}, \mathcal{I}) = \sum_{u=1}^{K+L} \sum_{\theta \in \Theta} \sum_{t_i=1}^{N_i} \sum_{f_i=1}^{\mathcal{F}_i} \mathcal{W}_s(\mathcal{X}, \mathcal{P}, \mathcal{I})$$
$$- \sum_{u=1}^{K+L} \sum_{\theta \in \Theta} \sum_{t_i=1}^{N_i} \sum_{f_i=1}^{\mathcal{F}_i} \tilde{\mathcal{Q}}_s(\mathcal{X}, \mathcal{P}, \mathcal{I}) \quad (27)$$

where $s \in \{E\}$ for $i = 1$, and $s \in \{E, L\}$ for $i = 2$.

Further, to convexify the constraint $\tilde{\mathcal{C}}_2$, we apply the first-order Taylor approximation on the second term of the equation. Now, the constraint $\tilde{\mathcal{C}}_2$ is re-written as

$$\overset{\triangle}{\tilde{\mathcal{C}}}_2 : \mathcal{Z}(\mathcal{X}) = \mu(\mathcal{X}) - (\nu(\mathcal{X}^{(j-1)}) + \nabla_\mathcal{X} \nu^T(\mathcal{X}^{j-1})(\mathcal{X} - \mathcal{X}^{(j-1)}) \leq 0, \quad (28)$$

where

$$\mu(\mathcal{X}) = \sum_{i \in \{1,2\}} \sum_{t_i=1}^{N_i} \sum_{f_i=1}^{\mathcal{F}_i} \sum_{u=1}^{K+L} x_{t_i, f_i}^u, \text{ and}$$

$$\nu(\mathcal{X}) = - \sum_{i \in \{1,2\}} \sum_{t_i=1}^{N_i} \sum_{f_i=1}^{\mathcal{F}_i} \sum_{u=1}^{K+L} (x_{t_i, f_i}^u)^2. \quad (29)$$

Later, to avoid the constraint $\overset{\triangle}{\tilde{\mathcal{C}}}_2$ from $\mathcal{OP}_3$, the objective function in $\mathcal{OP}_3$ is penalized and the optimization problem $\mathcal{OP}_3$ is reformulated as

$$\mathcal{OP}_4 : \max_{\{\mathcal{X}, \mathcal{P}, \mathcal{I}\}} \mathcal{R}''(\mathcal{X}, \mathcal{P}, \mathcal{I}) - q\mathcal{PC}''(\mathcal{X}, \mathcal{P}) - \lambda_1(\mu(\mathcal{X}) + \mathcal{Z}(\mathcal{X})) \quad (30)$$

subject to

$$\mathcal{C}_3, \mathcal{C}_4, \mathcal{C}_8, \text{ and } \mathcal{C}_9 \text{ as in } \mathcal{OP}_2,$$
$$\tilde{\mathcal{C}}_5, \tilde{\mathcal{C}}_6, \text{ and } \tilde{\mathcal{C}}_7$$

Now, the optimization problem $\mathcal{OP}_4$ is convex and can be solved [1] easily using standard convex optimization tools like CVX [38].

## V. NUMERICAL EVALUATIONS

In this section, we provide the simulation results to show the performance of the proposed joint allocation of the beam, RB, and transmit power technique for the slice-aware resource scheduling of the eMBB and URLLC users with the mixed and fixed numerologies in the OFDMA-based DL mmWave network. Importantly, we compare the performance of the proposed technique against the baseline techniques (i.e., (i) joint allocation of RBs of fixed numerology and beams and (ii) joint allocation of transmit power and RBs by assuming a single beam at the BS) in terms of achievable EE, total sum-rate, total power consumption, latency in delivered URLLC packets, and acquired data rates for eMBB users.

### A. Simulation Environment

In the considered mmWave DL network, a BS is located at the center of the cell coverage area with a radius of 150 m. eMBB and URLLC users are uniformly distributed across the entire network area. The channels between the BS and users are assumed to follow the extended Saleh-Valenzuela model. Further, all the simulation parameters related to the generation of channels are considered as in [28], [42]. The path loss exponents are considered 2 and 2.92 for Los and NLoS communication links, respectively. Each experiment is run for a time-frame duration of 10 ms, where the resource optimization is performed for every sub-frame duration of 1 ms by considering the actual queues of the users. However, according to the 3GPP standards, the beam pairing can be done for every $T_{ssb} \in \{5, 10, 20, 40, 80, 160\}$ ms [39], [40]. In this work, $T_{ssb}$ is fixed to 10 ms. Therefore, after identifying the aligned beam for every user in the first sub-frame of the time-frame using the CSI-RS, those beams are fixed for serving the aligned users in the remaining sub-frames. Furthermore, by following the 3GPP standards [41], we also consider the complete transmission bandwidth of 50 MHz and divide that into 66 or 33 RBs in every TTI according to the chosen numerology. The details of the considered resource grid models are provided as follows:

---

[1] Note that for the efficient performance of SCA-based algorithms, feasible initial points $(\mathcal{X}^{(0)}, \mathcal{P}^{(0)}, \mathcal{I}^{int})$ are required, which are obtained using the procedure given in [ [25], FIP1] for RBs and transmit power allocation. Also, each user is initially associated with the beam that provides the highest average SINR from all the RBs.

TABLE IV: Simulation Parameters

| Parameter | Value |
|---|---|
| Cell radius | 150m |
| Carrier frequency | 28 GHz |
| No. of eMBB users ($\mathcal{K}$) | 5, 10, and 15 |
| No. of URLLC users ($\mathcal{L}$) | 5, 10, 15, and 20 |
| No. of Beams at BS ($\mathcal{M}$) | 8 |
| Transmit power ($P_{max}$) | 50 dBm |
| LoS PL exponent | $\alpha_{los} = 2$ |
| LoS PL intercept | $\mathcal{A}_{los} = 10^{-6.41}$ |
| NLoS PL exponent | $\alpha_{nlos} = 2.92$ |
| NLoS PL intercept | $\mathcal{A}_{nlos} = 10^{-7.2}$ |
| Frame, Sub-frame lengths | 10 ms, 1 ms |
| Max. iterations ($T_{max}$) | 10 |
| Transmission bandwidth | 50 MHz |
| No. of antennas ($n_{T_x}$) | 8 per sub-array |
| Blocking parameter ($\beta$) | [0.003 0.02] |
| RB grid size for FN | $66 \times 8$ RBs with $\mu = 2$, $33 \times 16$ RBs for $\mu = 3$ |
| RB grid size for MN | $33 \times 8$ RBs with $\mu = 2$, $15 \times 16$ RBs with $\mu = 3$ |
| SCs, OFDM symbols per RB | 12, 7 |
| REs for transmission per RB | 60 |
| Guard band | 1910 KHz |
| URLLC traffic, packet size | FTP3 model, 32 bytes |
| URLLC packet size, $\lambda_u$ | 32 bytes, 4 packets/ms |
| eMBB traffic, packet size | Full-buffered, Infinite |
| Power used for SP per RB ($P_c$) | 0.005 |
| Static power consumption ($P_s$) | 0.05 |
| PA's drain efficiency ($\zeta$) | 0.25 |

*Fixed Numerology grid model:* In this model, all the available RBs in the time-frequency resource grid use the same numerology. Therefore, we use the numerologies with SCS 60 KHz and SCS 120 KHz for all the RBs in the grid. The complete grid within a sub-frame has $66 \times 8$ RBs, and $33 \times 16$ RBs with the numerology of SCS 60 KHz and SCS 120 KHz, respectively.

*Mixed Numerology grid model:* In this model, the effective transmission BW is divided into the two BWPs, where the BWP1 has RBs of numerology of SCS 60 KHz, and the BWP2 has RBs of numerology of SCS 120 KHz. Therefore, the complete grid within a sub-frame is comprised of $33 \times 8$ RBs with $\mu = 2$ and $15 \times 16$ RBs with $\mu = 3$. The total set of the simulation parameters is summarized in Table IV.

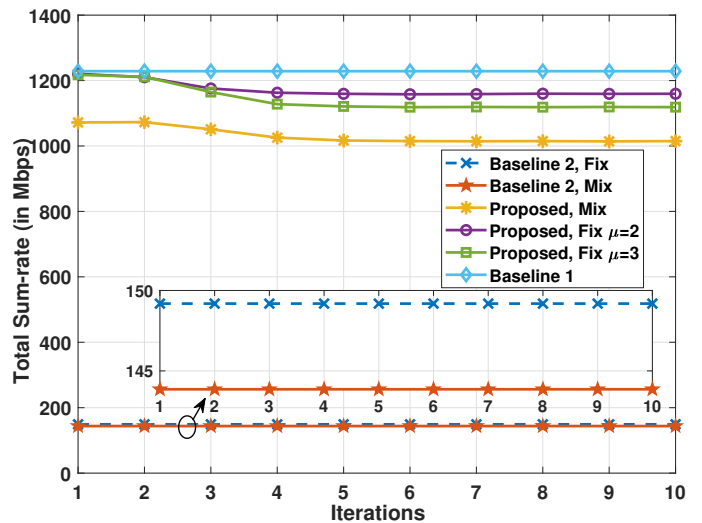

(a) Total sum-rate

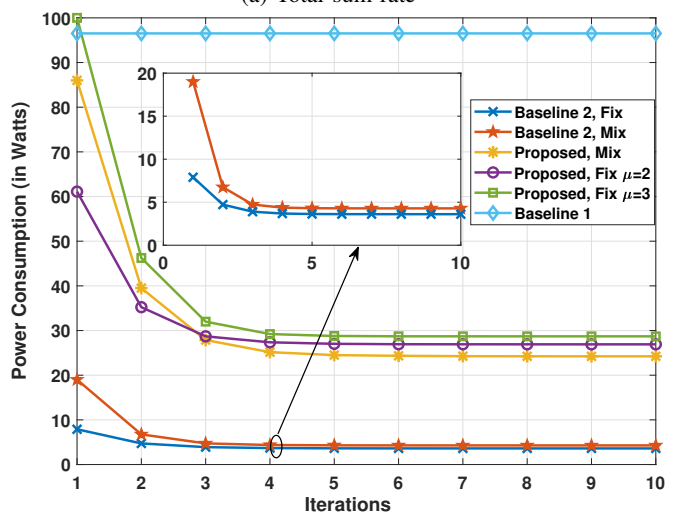

(b) Total power consumption

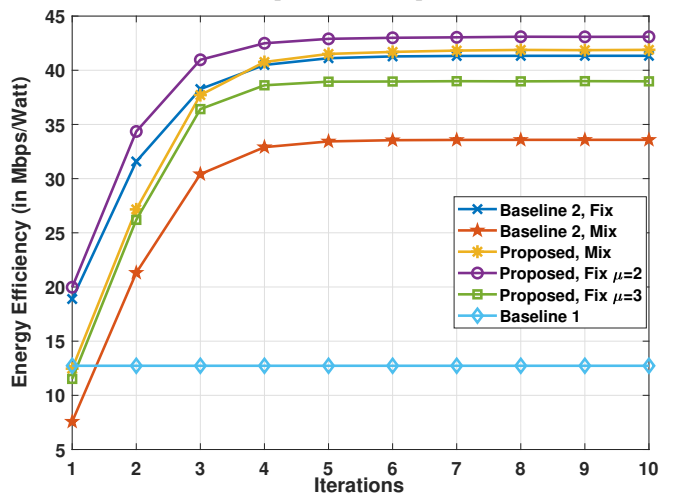

(c) EE performance

Fig. 4: Performance of the proposed and baseline methods with fixed and mixed resource grids at every iteration.

We compare the performance of the proposed method with the following two baseline methods.

**Baseline 1:** In this method, we assume the joint allocation of beams and RBs with fixed numerology of SCS 60 KHz (i.e., $\mu = 2$) for scheduling the users while considering the equally fixed transmit power to every available RB. Then, the optimization problem in (17) degrades to only the beams and RBs allocation, and it is solved using the proposed solution in Section IV.

**Baseline 2:** In this approach, we consider the allocation of RBs and transmit powers for scheduling the users using an omnidirectional antenna pattern at the BS. Then, this transmission power and RBs assignment problem is solved using a similar approach given in Section IV-C.

### B. Results and discussions

We first evaluate the performance of the proposed joint beams and resource allocation mechanism against the baseline approaches in terms of the achievable total sum rate, total power consumption, and EE.

We show the total achievable sum rate, total power consumption, and EE of the network for every iteration using the proposed and baseline approaches with fixed and mixed numerologies-based RBs grid models in Figs. 4(a)-4(c). From the results in Fig. 4(a), it is clear that the proposed method achieves a higher total sum rate than the total sum rate obtained using the baseline2 since the utilization of more beams exploits the spatial diversity and schedules more number of RBs to the users for data transmissions. Particularly, with the fixed numerology of SCS 60 kHz and SCS 120 kHz, the proposed method acquires a higher total sum rate compared to the acquired total sum rate with mixed numerology. The intuitive reason behind this result is that the mixed numerology wastes some RBs for guard bands to avoid the INI. As expected, baseline1 obtains a higher sum rate than the proposed method due to having equal high transmit powers on every RB.

As shown in Fig. 4(b), the power consumption of the network using the proposed and baseline1 methods is higher than the occurred power consumption using baseline2. This result occurs because the BS with multiple beams utilizes more power than a single beam (i.e., assumed in baseline2). Furthermore, the proposed method with fixed numerology of SCS 120 kHz consumes more transmit power than that with fixed numerology of SCS 60 kHz and mixed numerology. The decrement in SCS size increases the available RBs in every TTI of the RBs grid. For example, compared to SCS 120 kHz, the RBs grid with SCS 60 kHz has 33 more RBs in each TTI. In this case, the proposed scheduling method can leverage the frequency diversity to schedule the RBs with the best channel conditions for the users, leading to low power consumption. Likewise, with mixed numerology, as iterations increase, the proposed method utilizes lower transmit power consumption than the SCS 60 kHz and SCS 120 kHz by exploiting the frequency diversity in the RBs allocation to the users with low power consumption.

Furthermore, as can be seen from the results in Fig. 4(c), it is noticed that the proposed Dinkelbach-based iterative algorithm provides the improved solution set to acquire EE in every iteration until its convergence. Besides, it is observed from the results that the algorithm converges to the performance upper bound within 5 iterations for the considered system settings. Overall, we can assure from the simulation results that the proposed scheme fulfills the higher data rate requirement with the RBs grid model with the numerology of SCS 60 kHz at the cost of a small EE gap. As revealed later in this paper, the fixed numerology of SCS 60 kHz fails to guarantee the URLLC service's latency requirement. Precisely, in Figs. 4(a)-4(c), the results have shown the trade-off between the total achievable sum rate and EE of the network.

We show the performance of the proposed and baseline resource scheduling schemes in terms of latency during the delivery of URLLC packets in Fig. 5. Herein, we compute the overall latency in packet delivery as the addition of the packet waiting time in the queue, the RB scheduling time, and the data transmission time. As can be seen from the results in Fig. 5, it is evident that the proposed joint resource allocation and beam selection algorithm outperforms the baseline2 by obtaining a lower latency in the delivery of URLLC packets. In particular, the proposed algorithm achieves the lowest latency with the RB grid model based on mixed numerology. In addition, the results also show that the latency obtained in the delivery of packets decreases by decreasing the duration of the scheduling period. For example, with $T_f = 0.5$ ms, the proposed scheme using the mixed numerology grid delivers 100% of URLLC packets within 1 ms, while the proposed algorithm ensures the delivery of 80% and 59% of URLLC data packets within 1 ms with $T_f = 1$ ms using mixed and fixed numerology grids, respectively. Note that baseline1 shows the performance same as the proposed method with the fixed numerology of $\mu = 2$ at the cost of high energy consumption. Therefore, we avoid the baseline1 result here for brevity.

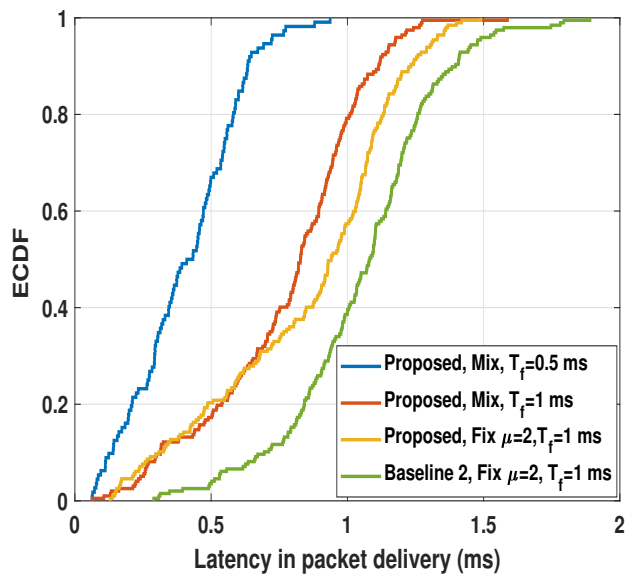

Fig. 5: ECDF of achievable latency in delivering packets using different RA mechanisms.

The empirical CDF (ECDF) of the achievable data rate for each eMBB user and the total sum rate of the eMBB users for varying numbers of total URLLC users using the proposed algorithm with MN and FN grid models and baseline 1 are shown in Fig. 7(a) and Fig. 7(b), respectively. It is observed from the results in Figs. 7(a), 7(b) that the proposed scheme obtains lower data rates for eMBB users with the increase in the number of eMBB and URLLC users. These results confirm that the proposed method dynamically schedules the resources for users according to their demands.

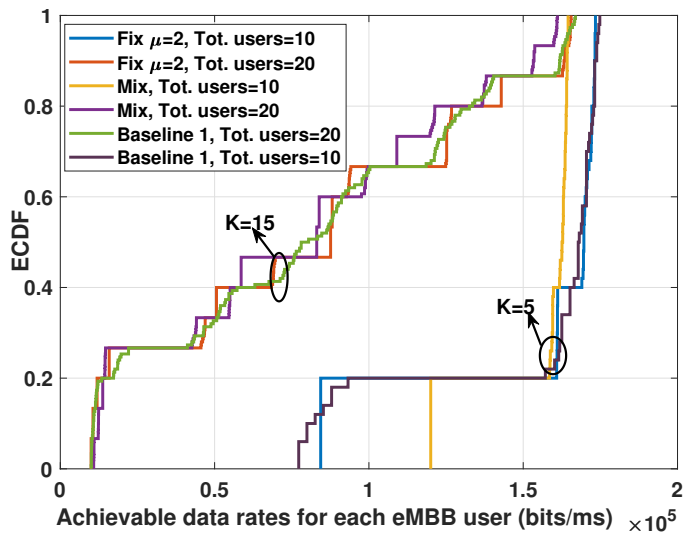

(a)

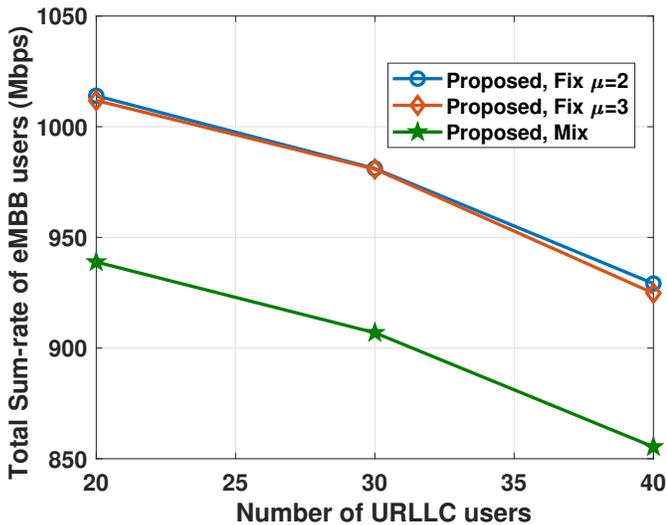

(b)

Fig. 6: (a) ECDF of achievable data rates for eMBB users and (b) total achievable sum-rate of eMBB users for varying numbers of URLLC users using the baseline and proposed methods with fixed and mixed numerology-based grid models.

The results are in Table. V portrays that the beams-selection algorithm dynamically schedules the beams according to the available active users in the network. For instance, when the active number of users in the cell is 30, the proposed beams selection scheme schedules almost 2 times a higher number of beams than that with the case having 5 users in the cell.

Next, we show the computational complexity of the proposed and baseline mechanisms in terms of the CPU run-time (in seconds) for each iteration in Fig. 7. These simulations are performed on a computer with a single CPU (i.e., Intel(R) Core(TM) i7-6820HQ) and developed using MATLAB 2018a with a CVX toolbox. The CPU run-time (i.e., the algorithm's execution time) is computed using the tic-toc command available in MATLAB programming. From the results in Fig. 7, it is observed that the proposed joint beams selection, RBs, and transmit powers allocation algorithm utilizes less computation time than the baseline2 method in the allocation of RBs to the users. Due to the utilization of more beams at the BS, each beam serves only the aligned users so that the scanning time is reduced to schedule the optimal RBs for users. Further, the baseline1 method (i.e., assigns only beams and RBs by considering the equal transmit power allocation) shows a lesser computation time than the other methods, as expected. It is also foreseen that because of having a high scanning time, the execution time of the proposed algorithm increases with the active number of users in the network. Furthermore, the results in Fig. 7 illustrate that the proposed algorithm uses more CPU time for allocating RBs with fixed numerology. When the fixed numerology-based RB's grid is available at the BS, the proposed algorithm scans the complete set of RBs for scheduling the eMBB and URLLC users, leading to high CPU run-time. On the contrary, the proposed algorithm scans only the available RBs in the assigned BWP to schedule either eMBB or URLLC users with optimal transmit powers when the mixed-numerology-based RB's grid is available at the BS. Overall, the BS utilizes less computation time for allocating RBs with mixed numerology using the proposed algorithm.

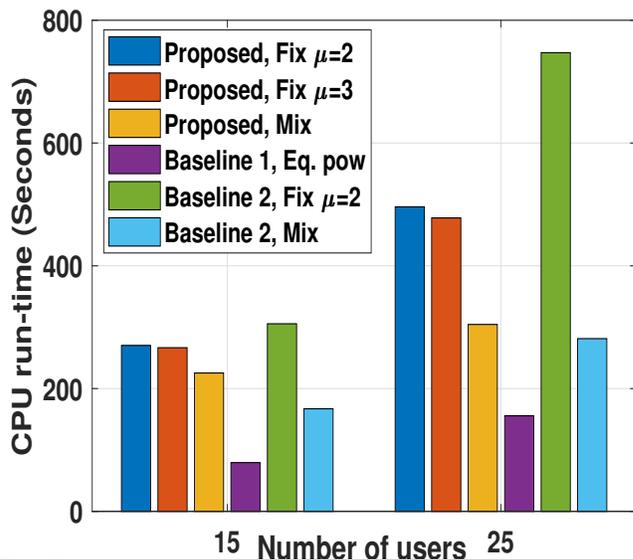

Fig. 7: Computational complexity of different RAN slicing methods.

## VI. CONCLUSIONS

In this paper, we investigated a joint resource assignment and beam selection problem for maximizing the overall network's EE while multiplexing the eMBB and URLLC users on the shared radio resources of an OFDMA-based DL mmWave network. This formulated optimization problem was identified as a MINLFP problem, which is intractable. To solve this, the formulated MINLFP problem was first transformed into a tractable MINLP problem (i.e., in the subtractive form) by leveraging the fractional programming theory. Subsequently, the transformed problem was solved by employing a two-loop Dinkelbach-based iterative solution. Importantly the resource allocation problem in the inner loop was solved using the DC programming and SCA method. Our simulation results illustrated that the proposed algorithm converges into the feasible solution within a less number of iterations and also demonstrated the trade-off between the achievable sum rate and EE for the proposed and baseline schemes with mixed and fixed resource grid models. Furthermore, the results also illustrated the impact of the blockages on the EE performance. Overall, the simulation results confirmed that the proposed method satisfies all users' requirements, including minimum data rate and packet latency with a mixed-numerology resource grid with lower computation time and at the cost of a moderate EE performance gap.

TABLE V: Beams utilization

| No. of users | 5 | 10 | 15 | 20 | 25 | 30 |
|---|---|---|---|---|---|---|
| No. of utilized beams | 3.88 | 5.90 | 6.92 | 7.44 | 7.70 | 7.85 |